\documentclass[preprint]{aastex}

\slugcomment{Accepted by the ApJL}

\shorttitle{The Origin of Solar Filament Plasma}

\shortauthors{Song et al.}

\begin{document}
\title{The Origin of Solar Filament Plasma Inferred from in situ Observations of Elemental Abundances}
\author{H.Q. SONG\altaffilmark{1,2}, Y. CHEN\altaffilmark{1}, B. LI\altaffilmark{1}, L.P. LI\altaffilmark{2}, L. ZHAO\altaffilmark{3}, J.S. HE\altaffilmark{4}, D. DUAN\altaffilmark{4}, X. CHENG\altaffilmark{5}, AND J. ZHANG\altaffilmark{6}}

\affil{1 Shandong Provincial Key Laboratory of Optical Astronomy
and Solar-Terrestrial Environment, and Institute of Space
Sciences, Shandong University, Weihai, Shandong 264209, China}
\email{hqsong@sdu.edu.cn}

\affil{2 Key Laboratory of Solar Activity, National Astronomical
Observatories, Chinese Academy of Sciences, Beijing 100012, China}

\affil{3 Department of Climate and Space sciences and Engineering,
University of Michigan, Ann Arbor, MI 48105, USA}

\affil{4 School of Earth and Space Sciences, Peking University,
Beijing 100871, China}

\affil{5 School of Astronomy and Space Science, Nanjing
University, Nanjing, Jiangsu 210093, China}

\affil{6 Department of Physics and Astronomy, George Mason
University, Fairfax, VA 22030, USA}

\begin{abstract}
Solar filaments/prominences are one of the most common features in
the corona, which may lead to energetic coronal mass ejections
(CMEs) and flares when they erupt. Filaments are about one hundred
times cooler and denser than the coronal material, and physical
understanding of their material origin remains controversial. Two
types of scenarios have been proposed: one argues that the
filament plasma is brought into the corona from photosphere or
chromosphere through a siphon or evaporation/injection process,
while the other suggests that the material condenses from the
surrounding coronal plasma due to thermal instability. The
elemental abundance analysis is a reasonable clue to constrain the
models, as the siphon or evaporation/injection model would predict
that the filament material abundances are close to the
photospheric or chromospheric ones, while the condensation model
should have coronal abundances. In this letter, we analyze the
elemental abundances of a magnetic cloud that contains the ejected
filament material. The corresponding filament eruption occurred on
1998 April 29, accompanying an M6.8 class soft X-ray flare located
at the heliographic coordinates S18E20 (NOAA 08210) and a fast
halo CME with the linear velocity of 1374 km s$^{-1}$ near the
Sun. We find that the abundance ratios of elements with low and
high First Ionization Potential such as Fe/O, Mg/O, and Si/O are
0.150, 0.050, and 0.070, respectively, approaching their
corresponding photospheric values 0.065, 0.081, and 0.066, which
does not support the coronal origin of the filament plasma.
\end{abstract}


\section{INTRODUCTION}

Solar filaments, also known as prominences when observed on the
solar limb, are one of the most common features in the corona. It
has been found that over 70\% of coronal mass ejections (CMEs) are
associated with filament eruptions (e.g., Webb \& Hundhausen 1987;
Gopalswamy et al. 2003). The associated CMEs usually exhibit
typical three-part structures, \textit{i.e.}, the bright leading
front, the dark cavity, and the bright core that is believed to
correspond with the filament (House et al. 1981). Filaments are
about one hundred times cooler ($T\sim$10$^{4}$ K) and denser than
the coronal material ($T\sim$10$^{6}$ K). Two crucial issues need
to be addressed to understand the filament formation (e.g.,
Parenti 2014; Li \& Zhang 2013; Zhou et al. 2014): what is their
magnetic structure (Yan et al. 2015, 2016) and where does their
material come from (Spicer et al. 1998; Mackay et al. 2010)? Our
understanding on the origin of the filament material remains
controversial. Two types of explanations have been proposed: one
argues that the filament plasma is brought into the corona from
photosphere and/or chromosphere. For instance, a siphon effect can
bring material into the corona from the photosphere or
chromosphere (e.g., Spicer et al. 1998), and the chromospheric
plasma can enter into the corona through evaporation or injection
process (Mackay et al. 2010; Parenti 2014); the other one suggests
that the surrounding coronal material condenses and forms the
filament due to thermal instability (e.g., Sakai et al. 1987;
D\'{e}moulin, 1993).

The elemental abundance measurement is a reasonable method to
constrain the models, because the coronal abundances are known to
differ from those of photosphere due to the first ionization
potential (FIP) effect (Schmelz et al. 2012; Laming 2015).
Usually, the abundance ratios of elements with low FIP ($<$10 eV,
\textit{e.g.}, iron, magnesium, and silicon) and those with high
FIP ($>$10 eV, \textit{e.g.}, oxygen and neon) will be enhanced in
the corona, which is supported by the spectroscopic and solar wind
data (e.g., Schmelz et al. 2012 and references therein). For
example, it is generally accepted that magnetic flux ropes (MFRs)
form in the corona, therefore large FIP fractionations are
frequently associated with magnetic clouds (MCs, Zurbuchen et al.
2004; Smith et al. 2001). This elemental fractionation results
from a separation of ions and neutrals, taking place between the
photosphere and corona. The siphon and evaporation/injection model
would predict that the filament material abundances are close to
the photospheric or chromospheric ones, while the condensation
model should have coronal abundances. Some abundance analyses
through spectroscopy have been conducted to address this issue
(Widing et al. 1986; Spicer et al. 1998). Their results showed
that the magnesium-to-neon ratios have values intermediate between
the photosphere and corona but none are as high as expected in the
corona. This indicates the filament material should not originate
from the corona (Spicer et al. 1998).

Spectroscopy method has been pretty good and widely used to
determine the abundances of astrophysical objects, while the
method has a particular problem for the filament analysis because
of its optical thickness. The radiation transfer process will
hinder us to infer the precise elemental abundances of its
interior. Therefore, it is still an open question about the origin
of filament material. An alternative method to address this issue
is to measure the filament abundances in the interplanetary space,
which requires a clear identification of the filament material
with in situ data. The filament is cooler in the corona as
mentioned, so filament plasma will yield a significant fraction of
ions with low-ionization temperatures, \textit{e.g.}, Fe$^{6+}$,
instead of Fe$^{10+}$ in background solar wind or Fe$^{16+}$ in
hot regions within ICMEs (Song et al. 2016). This can be used to
identify the filament in the interplanetary space (Lepri \&
Zurbuchen, 2010).

As mentioned, over 70\% of CMEs are associated with filament
eruptions. However, only $\sim$4\% of detected ICMEs have
significant contributions of ions with low charge states (Fe$^{4+}
\sim $Fe$^{7+}$), \textit{i.e.}, exhibiting the signature of
filaments (Lepri \& Zurbuchen 2010), and usually their filament
durations last only 1 to several hours (Yao et al. 2010), too
short for the elemental abundance analysis as the temporal
resolution is 2 hours. In an earlier statistical study (Song et
al. 2016), we found a filament eruption event on 1998 April 29
that had a long ($\sim$20 hours) duration when propagating to the
L1 point. Gloeckler et al. (1999) found anomalously large
enrichment of $^{3}$He$^{++}$/$^{4}$He$^{++}$, He/O and Fe/O in
this event. Skoug et al. (1999) reported that there existed a
prolonged He$^{+}$ enhancement within this ICME, indicating the
prominence material. However, neither of them paid attention to
the origin of filament plasma. In this letter, we revisit this
event to address this issue through investigating the filament
elemental abundances. \textit{Advanced Composition Explorer (ACE)}
provides both the Fe charge state and elemental abundance
information, which is used to identify the filament material and
infer its origin. The proton velocity distribution functions
(VDFs, Marsch et al. 1982; He et al. 2015) inside the filament are
also analyzed. We introduce the instruments and methods in Section
2 and present the observational results in Section 3. Our summary
and discussion is presented in the final Section.

\section{INSTRUMENTS AND METHODS}

The eruption process is recorded by the Extreme Ultraviolet
Telescope (EIT, Delaboudiniere et al. 1995) 195 \AA\ passband and
Large Angle and Spectrometric Coronagraph (LASCO, Brueckner et al.
1995) on board the \textit{Solar and Heliosphere Observatory
(SOHO)}. With the EIT 304 \AA\ data, we identify that this event
was associated with a filament eruption. The soft X-ray (SXR) data
are from the \textit{Geostationary Operational Environment
Satellite (GOES)}, which provides the integrated full-disk SXR
emission from the Sun.

The \textit{in situ} ion-charge states and elemental abundances
are from the Solar Wind Ion Composition Spectrometer (SWICS,
Gloeckler et al. 1998) on board the \textit{ACE}, which was
launched in 1997 and orbiting around the L1 point. As a mass
spectrometer, SWICS consists of a time-of-flight system (TOF) and
an energy resolving solid-state detector (SSD). Through measuring
the speed of each ion and determining its residual energy with TOF
and SSD, respectively, the mass, charge, and energy of each
detected ion can be identified. SWICS can provide the charge-state
distributions and abundances of $\sim$10 elements (Lepri et al.
2001). Here the newly released SWICS 1.1 level 2 data are used
(Shearer et al. 2014), which have a better accuracy than the
earlier versions with improved calibration. The new version
improved the process to remove accidental coincidence events,
resulting in an increasing identification of rare ions
(\textit{e.g.}, Fe$^{6+}$ and Fe$^{7+}$). This is crucial to
identify the filament and to diagnose its abundances in the
present study. We also use the magnetic field and plasma data from
MAG (Smith et al. 1998) and SWEPAM (McComas et al. 1998) on board
\textit{ACE}. The proton three-dimensional (3D) VDFs within the
ICME are measured by 3DP (Lin et al. 1995) onboard \textit{WIND}.
Here we adopt a tri-quadratic interpolation method (Marsch et al.
1982; He et al. 2015) to reconstruct the full 3D VDF from the 3DP
data.

\section{OBSERVATIONS AND RESULTS}
\subsection{Overview of the Eruption}
An M6.8 class SXR flare, located at the heliographic coordinates
S18E20 (NOAA 08210) from the Earth perspective, was recorded by
\textit{GOES} on 1998 April 29. The corresponding SXR flux rose
from $\sim$16:06 UT and peaked at $\sim$16:37 UT as shown in
Figure 1(a). The eruption process was observed by the EIT 195 \AA\
($\sim$1.5 MK) passband as exhibited with the animation available
online. We present its post-eruptive arcades (PEAs, Tripathi et
al. 2004) in Figure 1(b), and the associated CME recorded by the
LASCO C3 with field of view (FOV) of 4--18 R$_\odot$ in Figure
1(c). The CME was a halo one due to its propagation toward the
Earth with a linear velocity of 1374 km s$^{-1}$ in the LASCO FOV
(https://cdaw.gsfc.nasa.gov/). To make sure that this is a
filament eruption event, we checked the EIT 304 \AA\ ($\sim$0.06
MK) images, which are suitable to identify the cooler filament.
There were only two frames on 1998 April 29 for EIT 304 \AA.
Fortunately, these two frames were taken before and after the
eruption respectively and confirmed that the event was a filament
eruption (see Figures 1(d) and (e)). The green arrows and the
dotted line in Figure 1(d) depict the filament prior to the
eruption. In Figure 1(e), the filament was invisible due to its
full eruption.

According to the online ICME list \\
 (http://www.srl.caltech.edu/ACE/ASC/DATA/level3/icmetable2.htm)
compiled by Richardson and Cane (Cane \& Richardson, 2003;
Richardson \& Cane, 2010), the shock generated by this ICME
arrived at \textit{ACE} on 1998 May 1. We also evaluate the
arrival time with the shock propagation model (SPM, Feng \& Zhao,
2006; Zhao \& Feng 2014), which is based on an analytic solution
of blast wave to predict the interplanetary shock arrival times.
Here the SPM model predicts that the shock should arrive around
21:00 UT on 1998 May 1, very close to the observed moment at 21:21
UT as shown in Figure 2. Therefore, there is no doubt that the
ICME detected by \textit{ACE} corresponds to the halo CME
associated with the filament eruption.

\subsection{The Filament Plasma and Elemental Abundance Analysis}

In 1998 May, both \textit{ACE} and \textit{WIND} orbited around
the L1 point, and the distance between them can be neglected
compared with the much larger ICME scale (see e.g., Song et al.
2016). We demonstrate this point in Figures 2(a) and (b), which
show the magnetic filed measured by \textit{ACE} and
\textit{WIND}, respectively. The total magnetic field strength is
presented with black lines and its three components are plotted in
the Geocentric Solar Ecliptic coordinate. The X-axis (red) points
from the Earth toward the Sun, the Y-axis (green) points toward
the opposite direction of the Earth's motion, and the Z-axis
(blue) is parallel to the ecliptic pole. It is obvious that the
magnetic field profiles in these two panels are almost identical,
proving that both spacecraft passed through the same part of the
ICME structure. The shock, pointed out with the purple vertical
solid line, arrived at 21:21 UT on May 1, and the magnetic cloud
(MC), exhibiting low thermal velocities of protons, started from
11:31 UT on May 2 and ceased until 16:48 UT on May 3. The MC
boundaries are marked with two purple vertical dot-dashed lines.

Figure 2(c) is the normalized pitch angle distribution (PAD) of
272 eV electrons measured by \textit{ACE}, exhibiting
bidirectional electrons (BDEs) during the MC passage as expected
if the MFR two footpoints keep anchored on the Sun during the
propagation. The velocity profile within the MC decreases, from
650 to 430 km s$^{-1}$, with time as shown in Figure 3(d),
indicating the expansion of the MFR. Figure 3(e) presents the
thermal velocity (red) and number density (black) of protons. One
of the typical MC characteristics, \textit{i.e.}, lower
temperature, is demonstrated clearly with the red solid line. The
thermal velocity within MC ($\sim$15 km s$^{-1}$) corresponds to
$\sim$10$^{4}$ K. While one interesting phenomenon is that there
exists an extended region of high density, over 30 cm$^{-3}$ in
general, as emphasized with yellow shade, corresponding to the
presumed filament material. The dynamical and total perpendicular
pressures are plotted in Figure 2(f) with red and black lines,
respectively, showing that the pressure within the shade
considerably larger than those of the other regions.
Correspondingly, the plasma $\beta$ (red line in Figure 2(g))
within the shade is greatly enhanced reaching an maximum of 10,
due to the high density there. This is not the typical MC
characteristics (Burlaga et al. 1981). The black line in Figure
2(g) demonstrates that the entropy of the shaded region is lower
than that of the other regions. Note Figures 2(e)--(g) are
provided by \textit{WIND} as \textit{ACE} does not have the
corresponding data within the shade.

The Fe charge state distributions with a temporal resolution of 2
h, ranging between 6$+$ and 20$+$, in Figure 2(h) show that large
amount of low ionization states such as 6$+$ and 7$+$ exists in
the yellow shade. Correspondingly, the average iron charge state
($<$Q$>$Fe) decreased from over 16$+$ to near 6$+$ as demonstrated
with the white solid line, which is obviously lower than those of
the solar wind ($\sim$ 9$+$ to 11$+$). Note the white horizontal
dotted line represents the position of 9$+$. All of the observed
features, including the low temperature, high density, and low Fe
charge states, are consistent with the filament characteristics,
providing strong evidence that the shaded region is the
interplanetary counterpart of the erupted filament as presented in
Figure 1 (see also Lepri \& Zurbuchen, 2010).

Figure 3 shows the elemental abundances to infer the origin of the
filament material with SWICS. The Fe charge state distributions
are re-plotted in Figure 3(a) to identify the filament interval.
For comparison, two regions with Fe charge states higher and lower
than those in the normal solar wind are emphasized with yellow
shades, which are the non-filament (left) and filament (right)
intervals within the ICME, respectively. Note that to represent
the filament, we select a shorter region in Figure 3 (in
comparison with Figure 2) with dominated lower Fe charge states
($<$Q$>$Fe $<$ 9$+$).

As mentioned, the abundance ratios of elements with low and high
FIP (e.g., Fe/O, Mg/O, and Si/O) will be enhanced in the corona,
and large FIP fractionations are frequently associated with MCs as
they form in the corona. The variations of abundance ratios within
MCs indicate their plasmas should have different origins. Figures
3(b)-(d) show the Fe/O, Mg/O, and Si/O ratios sequentially with a
temporal resolution of 2 h. The horizontal dotted lines in these
panels depict the corresponding elemental ratios of the solar
photosphere (Asplund et al. 2009). The ratios in the non-filament
region are obviously different from those in the filament region.
For instance, the Fe/O (Mg/O and Si/O) ratios in non-filament
region are around 0.25 (0.30 and 0.30), while those in the
filament region approach 0.150 (0.050 and 0.070), which are closer
to the photospheric values of 0.065 (0.081 and 0.066). Therefore,
our in situ observations do not support that the filament material
originates from the corona.

\subsection{The Velocity Distribution Functions within the Filament}
Yao et al. (2010) presented in situ observations of three MCs that
contain cold and high-density material at 0.3, 0.5, and 0.7 AU
with \textit{Helios 2}. They investigated the microstates of the
filaments with the thermal proton VDFs, which were cooler compared
to non-filament plasmas. But two filaments showed obvious
anisotropic feature, and the other one was close to isotropic. It
is likely that different features originated from different
filaments. However, the intervals with filament in their events
were very short ($\sim$1 h), so they did not examine whether the
difference of VDFs exists within the same filament.

Here the interval of filament lasted about 20 h, allowing us to
examine the VDFs at different locations with \textit{WIND}
measurements. Figure 4 illustrates the cross-section view of the
VDFs from one orthogonal visual angle at different locations as
marked with filled circles in Figures 2(e) and 3(a). The cut view
here is defined by the coordinate axis pair [$R$,
($B_{0}$$\times$$R$)$\times$$R$], where $R$ is the radial
direction and $B_{0}$ is the local mean magnetic field vector.
Figure 4(a) shows the non-filament region, and (b)-(d), the
filament. It is obvious that the VDFs in (a) is larger compared to
those presented in other panels, which is consistent with their
respective temperatures. The selected non-filament region
approaches isotropic, yet different types of VDFs appear within
the filament. Figure 4(b) exhibits an isotropic VDF, while (c) and
(d) demonstrate a dominant heating perpendicular to the magnetic
field indicated by the white solid line. Therefore, we confirm
that both isotropic and anisotropic features can appear within the
same one filament. However, it is not clear yet that the observed
VDF difference originates in the solar atmosphere or is formed
during the filament propagation in the interplanetary space.

\section{SUMMARY AND DISCUSSION}
In this letter, an erupted active-region filament, accompanied by
an M6.8 class SXR flare and a halo CME with linear velocity of
1374 km s$^{-1}$ near the Sun (https://cdaw.gsfc.nasa.gov/), was
detected by both \textit{ACE} and \textit{WIND} at the L1 point,
which was unambiguously proved by the high density, low
temperature, and low Fe charge states in the interval. The
elemental ratios with low and high FIP were deduced to infer the
origin of filament material. The results show that the Fe/O (Mg/O
and Si/O) ratios of non-filament interval are around 0.25 (0.30
and 0.30), while those of filament interval approach 0.150 (0.050
and 0.070), closer to the photospheric values of 0.065 (0.081 and
0.066). As this filament originated from an active region (NOAA
08210), and its origin may be different from that of quiescent
filaments. Therefore, our in situ observations do not support that
active-region filament plasma originates from the corona. The
analysis of VDFs demonstrates that the same filament can exhibit
both isotropic and anisotropic features near 1 AU.

As mentioned, only $\sim$4\% of detected ICMEs exhibit the
signature of filaments and most of them have a short duration of 1
to several hours. This may be partially attributed to that the
filament material only occupies a small fraction of ICME. The
filament reported in this study is very special and prominent for
its extended duration ($\sim$20 h). We use a velocity-modified
Gold and Hoyle model to fit the MC (Gold \& Hoyle, 1960; Wang et
al. 2016), and find the spacecraft passed through the interior of
the MC, close to the MC center. This may explain why the filament
interval was so long. Nevertheless, this event offers us an
excellent opportunity to investigate the elemental abundances of
filament interior with in situ data.

Given that the FIP fractionation takes place between the
photosphere and corona, the elemental abundances at different
altitudes in the chromosphere are expected to vary between the
photospheric and coronal values. Therefore, our analysis cannot
definitively determine whether the filament plasma comes from the
photosphere or chromosphere. Nonetheless, in agreement with Spicer
et al. (1998), our study does not support the coronal origin of
filament plasma.

As mentioned, large FIP fractionations (e.g., high Fe/O ratios)
are frequently associated with MCs. However, within some MCs,
there exist low Fe/O ratios close to the photospheric values
without accompanying filament features (low Fe charge state and
high proton density). These plasmas might correspond to the
chromospheric ones that are evaporated into the corona along the
magnetic field lines stretched by the rising MFRs during the
eruptions and then involved into the MFRs through magnetic
reconnection. The evaporated plasmas during eruptions are hot with
high charge states and low Fe/O ratios, which are different from
the filament material with low charge states and low Fe/O ratios.

\acknowledgments We are grateful to the referee, Drs. S.T. Lepri,
Y.M. Wang, P.F. Chen, J. Lin, S. Yao, H. Tian, X.L. Yan, X.H.
Zhao, G. Li, and Q. Hu for their valuable comments and discussion.
We acknowledge the use of data from the \textit{GOES},
\textit{SOHO}, \textit{ACE} and \textit{WIND} missions. This work
is supported by NSFC grants 41274177, 41274175, and 41331068. J.Z.
is supported by US NSF AGS-1249270 and NSF AGS-1156120. L.Z. is
supported by NSF grants AGS-1344835 and AGS-1621686. H.Q.S. is
also supported by the open research program of CAS Key Laboratory
of Solar Activity.

\clearpage

\begin{figure}
\epsscale{0.85} \plotone{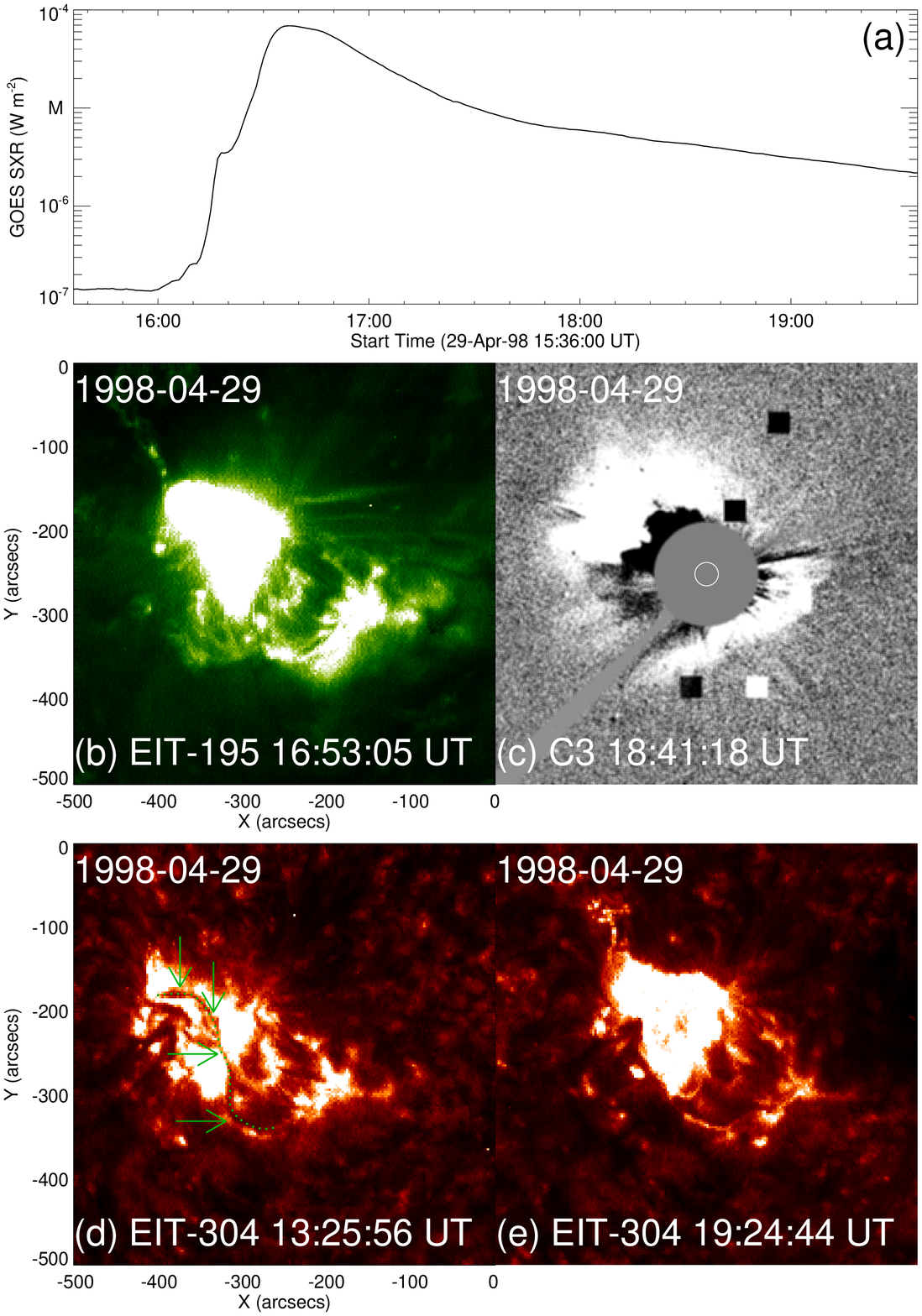} \caption{A filament eruption
event on 1998 April 29. (a) The \textit{GOES} SXR 1-8 \AA \ flux
profile of the accompanying flare. (b) EIT 195 \AA\ ($\sim$1.5 MK)
image after the eruption, showing the PEAs at high temperature.
(c) The associated halo CME recorded by LASCO C3. (d)--(e) The EIT
304 \AA\ ($\sim$0.06 MK) images, showing the low-temperature
filament visible (prior to eruption) and invisible (after
eruption), respectively. (An animation of this figure is
available.) \label{Figure 1}}
\end{figure}

\begin{figure}
\epsscale{0.80} \plotone{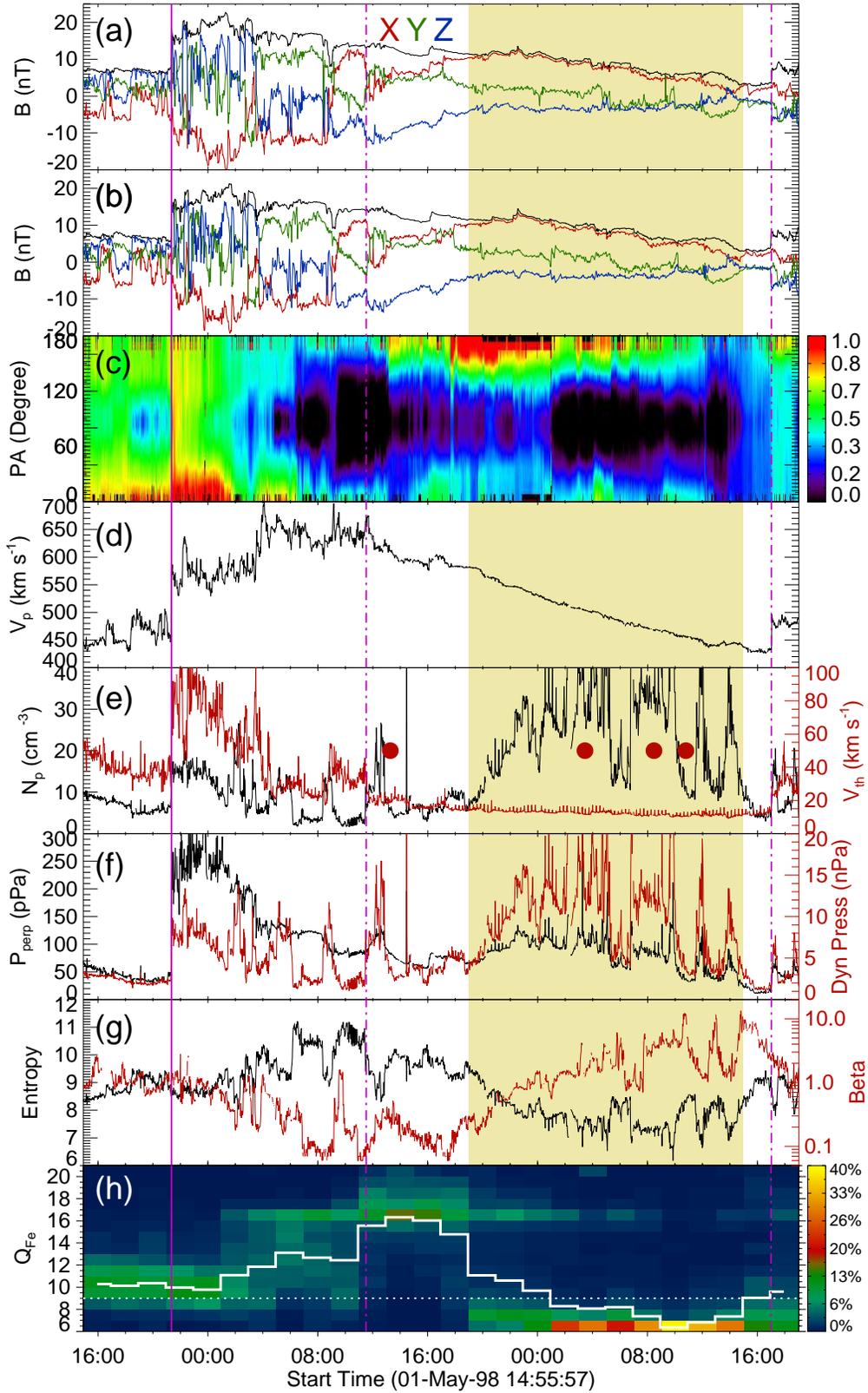}
 \caption{Plasma parameters measured with \textit{ACE} and \textit{WIND}.
 The panels show the total magnetic field strength (black) and its X (red), Y (green) and Z (blue) components from both \textit{ACE} and \textit{WIND} (a and b), PAD of electrons at 272 eV (c), bulk speed (d),
 thermal velocity (red) and density (black) (e), the dynamic pressure (red) and the total pressure along the perpendicular direction (black) (f), plasma $\beta$ (red) and entropy (black) (g),
 Fe charge-state-distribution map, as well as the $<$Q$>$Fe (h). The purple vertical solid line represents the shock and the dot-dashed lines mark the MC boundaries.
\label{Figure 2}}
\end{figure}

\begin{figure}
\epsscale{0.85} \plotone{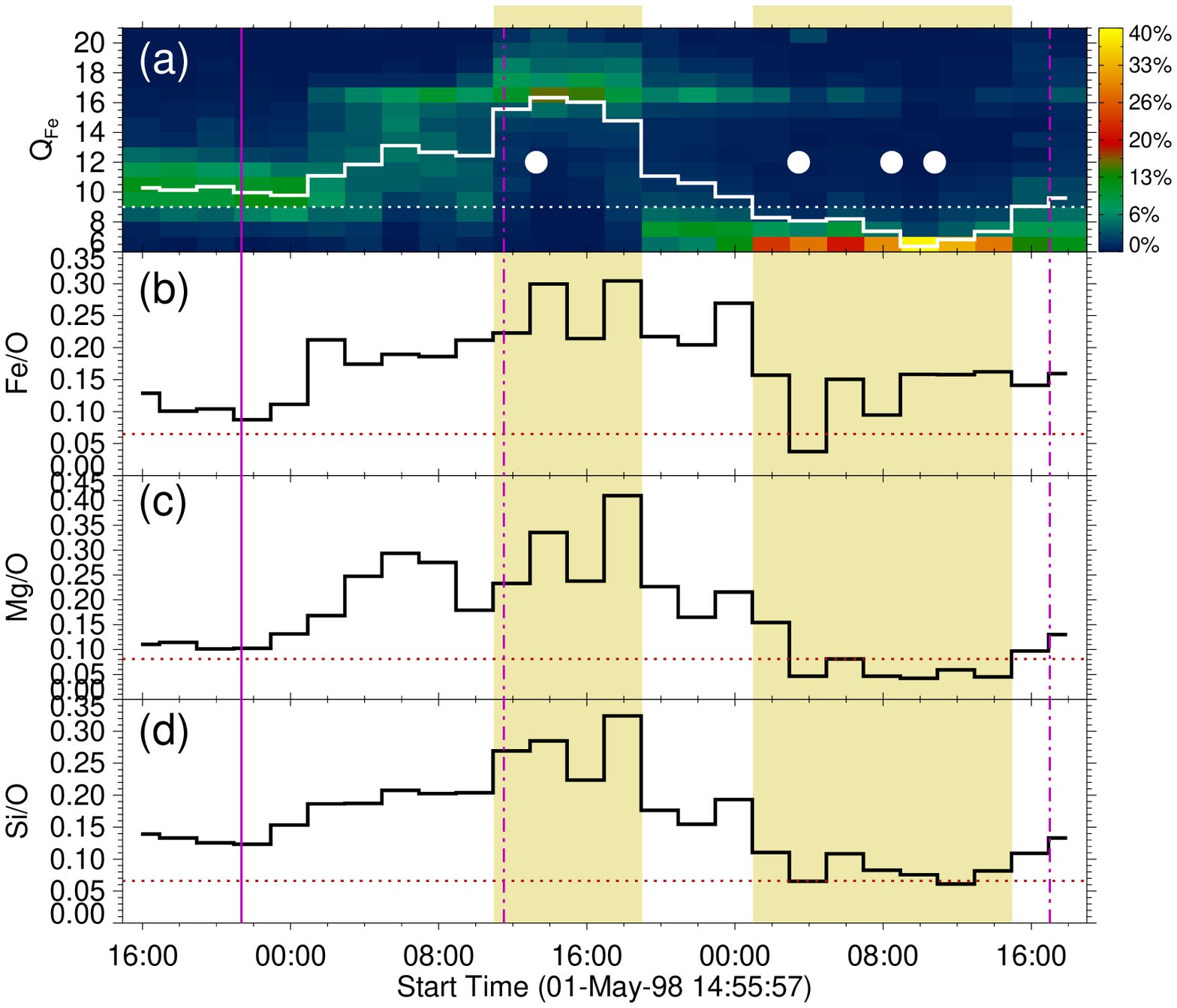} \caption{Charge states and
elemental abundances measured with \textit{ACE}. The panels show
the Fe charge-state-distribution map and the $<$Q$>$Fe profile
(a), the temporal profiles of Fe/O (b), Mg/O (c), and Si/O (d).
The red horizontal dotted lines in panels (b)-(d) shows the
corresponding elemental ratios of the solar photosphere.
\label{Figure 3}}
\end{figure}

\begin{figure}
\epsscale{0.4} \plotone{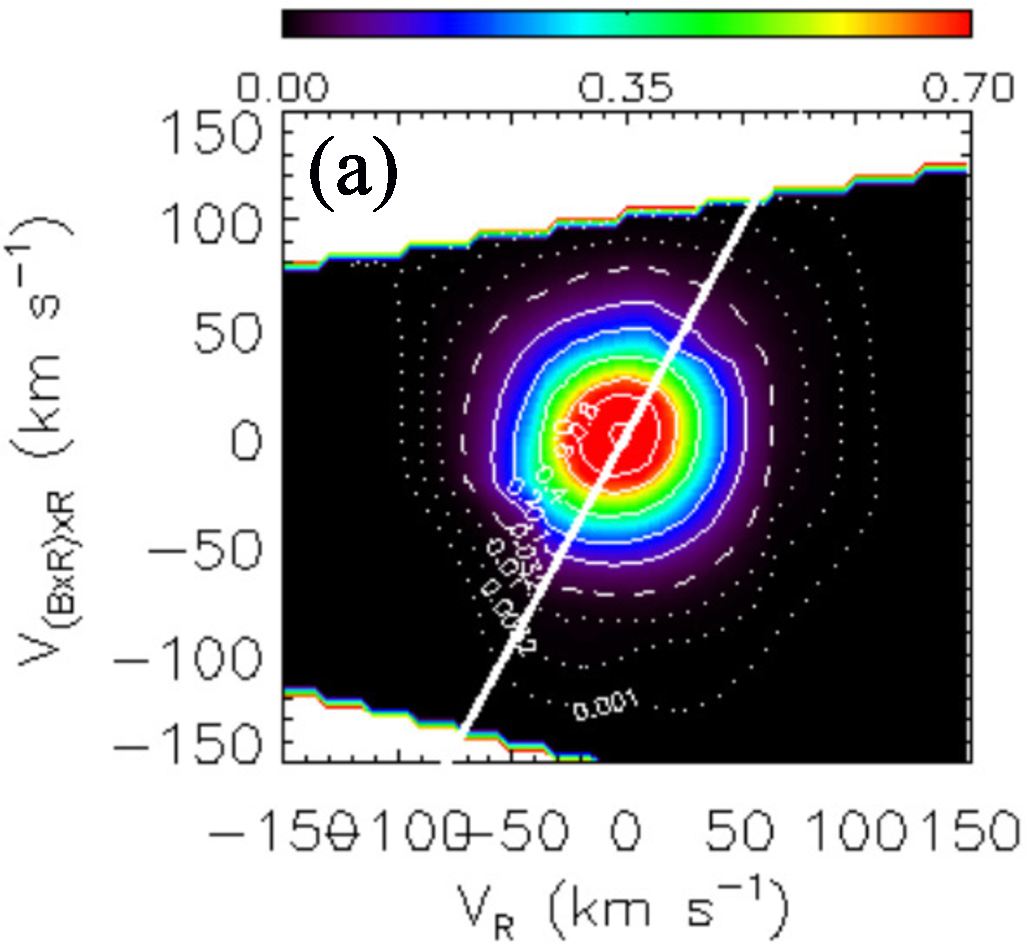} \plotone{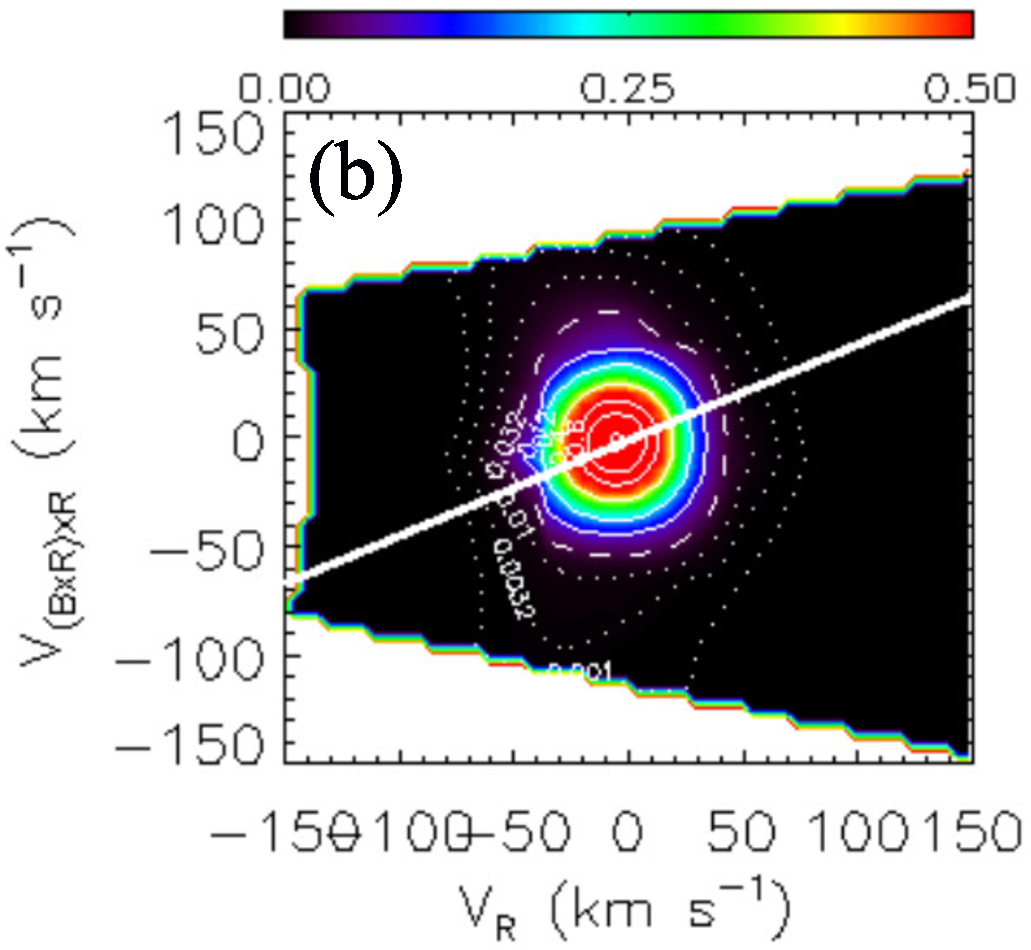}
\plotone{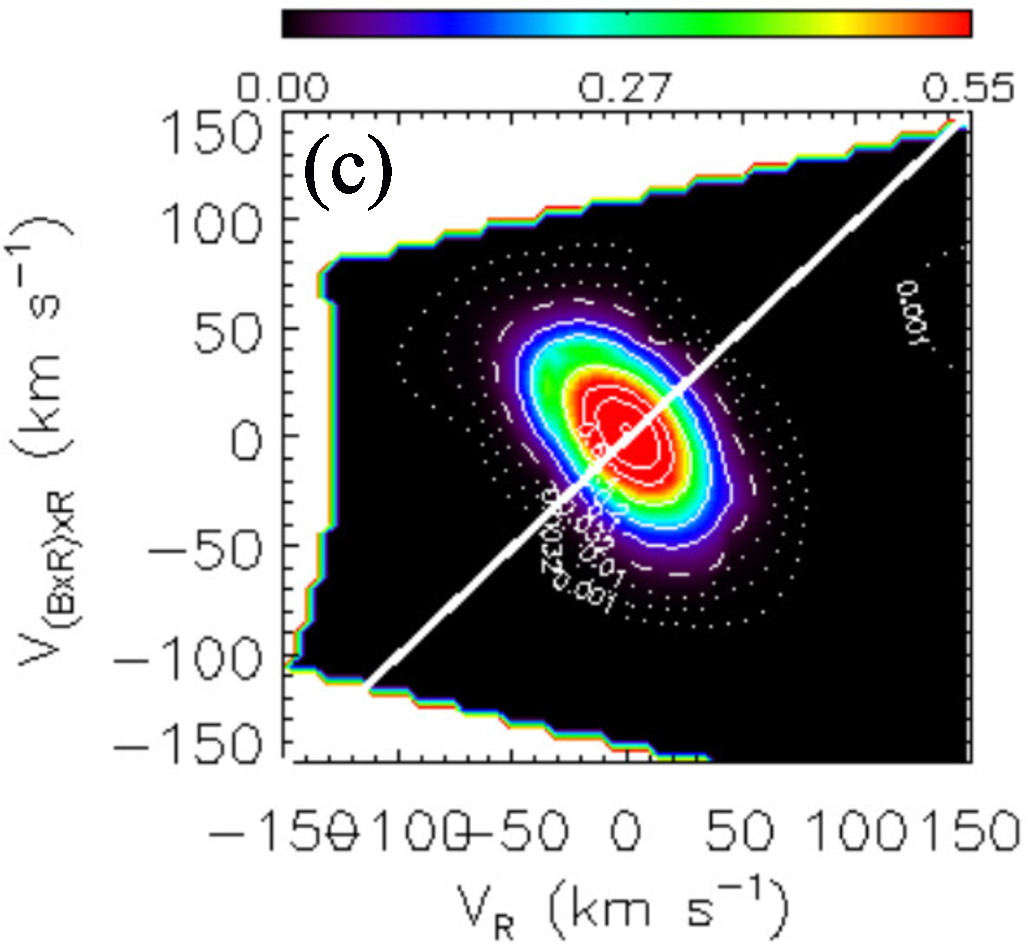} \plotone{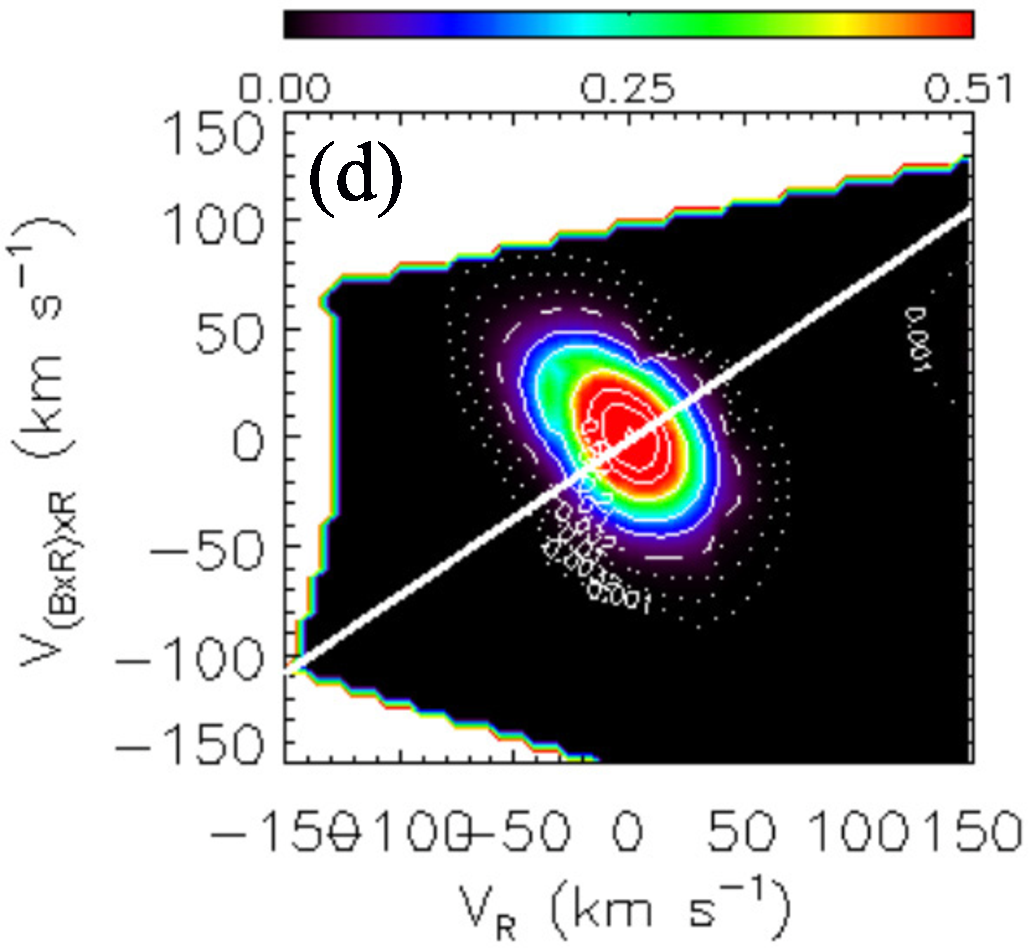} \caption{Cross sections of
VDFs measured with \textit{WIND}. Panels (a)-(d) correspond to the
positions marked with four filled circles in Figures 2(e) and
3(a). The coordinate origin is set at the velocity of the maximum
reduced-VDF values. The contours are 0.98, 0.8, 0.6, 0.4, 0.2,
0.1, 0.032, and 0.001 from inner to outer sequentially. The white
solid line indicates the direction of background magnetic field.
\label{Figure 4}}
\end{figure}

\end{document}